\documentclass{emulateapj}
\usepackage{graphicx}
\usepackage{color}

\newcommand{\beq}	{\begin{equation}}
\newcommand{\eeq}	{\end{equation}}
\newcommand{\beqa}{\begin{eqnarray}}
\newcommand{\eeqa}{\end{eqnarray}}

\newcommand{\e}	{$^{-1}$}
\newcommand{\ee}	{$^{-2}$}
\newcommand{\eee}	{$^{-3}$}
\def\simlt{\lower.5ex\hbox{$\; \buildrel < \over \sim \;$}}
\def\simgt{\lower.5ex\hbox{$\; \buildrel > \over \sim \;$}}
\def\la{\simlt}
\def\ga{\simgt}
 
\font\tenbi=cmmib10 
\newfam\bifam \def\bi{\fam\bifam\tenbi} \textfont\bifam=\tenbi

\font\tenbr=cmbx10
\newfam\brfam \def\br{\fam\brfam\tenbr} \textfont\brfam=\tenbr

\font\squinttenbi=cmbx10 at 9pt
\scriptfont\brfam=\squinttenbi

\def\vecnabla{
              \setbox1=\hbox{$\bigtriangledown$}
                           \raise.45ex\hbox{$\bigtriangledown$\hskip-.97\wd1
                           $\bigtriangledown$\hskip-.97\wd1
                           $\bigtriangledown$\hskip-.97\wd1}
                           \raise.47ex\hbox{$\bigtriangledown$}}
\def\grad{\vecnabla}

\def\div{\vecnabla\cdot}

\def\vece{{\bi{e}}}

\def\vecF{{\bi{F}}}
\def\rvecP{{\br{P}}}

\newcommand{\caln}		{{\cal N}}
\newcommand{\calr}		{{\cal R}}

\newcommand{\fdiss}	{f_{{\rm diss}}}
\newcommand{\fedd}		{f_{\rm Edd}}
\newcommand{\esh}		{\hat E^*}
\newcommand{\esho}	{E^*_1}
\newcommand{\eosh}	{\hat E_0^*}
\newcommand{\fsh}		{\hat F^*}
\newcommand{\fshield}	{f_{\rm shield}}
\newcommand{\fsho}		{F^*_1}
\newcommand{\fcs}		{F^*_{\rm ch}}
\newcommand{\hone}	{{\rm H\,I}}
\newcommand{\honet}	{{\rm H\,I, tot}}
\newcommand{\hones}	{{\rm H\,I,\, slab}}
\newcommand{\fone}		{f_1}
\newcommand{\htwo}	{{H$_2$}}
\newcommand{\mhtwo}	{{{\rm H}_2}}
\newcommand{\ins}		{I_\nu^*}
\newcommand{\lch}		{\ell_{\rm ch}}
\newcommand{\muh}		{\mu_{\rm H}}
\newcommand{\nh}		{n_{\rm H}}
\newcommand{\phik}		{\phi_\kappa}
\newcommand{\stn}		{\sigma_{2\nu}}
\newcommand{\strom}	{Str{\"o}mgren}
\newcommand{\inc}		{{\rm inc}}
\def\veceh			{{\hat\vece}}

\shorttitle{The Atomic-to-Molecular Transition in Galaxies III}
\shortauthors{McKee \& Krumholz}
\begin{document}
\title{The Atomic-to-Molecular Transition in Galaxies. III. A New Method for Determining the Molecular Content of Primordial and Dusty Clouds}
\author{Christopher F. McKee}
\affil{Physics Department and Astronomy Department, University of California,
    Berkeley, CA 94720}
\email{cmckee@astro.berkeley.edu}
\and
\author{Mark R. Krumholz}
\affil{Astronomy Department, University of California,
    Santa Cruz, CA 95060}
\email{krumholz@ucolick.edu}

\begin{abstract}
Understanding the molecular content of galaxies is a critical problem in star formation and galactic evolution. Here we present a new method, based on a \strom-type analysis, to calculate the amount of \hone\ that surrounds a molecular cloud irradiated by an isotropic radiation field. We consider both planar and spherical clouds, and H$_2$ formation either in the gas phase or catalyzed by dust grains. Under the assumption that the transition from atomic to molecular gas is sharp, our method gives the solution without any reference to the photodissociation cross section. We test our results for the planar case against those of a PDR code, and find typical accuracies of about 10\%. Our results are also consistent with the scaling relations found in Paper I of this series, but they apply to a wider range of physical conditions. We present simple, accurate analytic fits to our results that are suitable for comparison to observations and to implementation in numerical and semi-analytic models.
\end{abstract}

\section{Introduction}

\subsection{Description of the Problem}

Stars form in molecular gas. It is therefore of central importance to understand how
molecular gas forms in the interstellar medium of galaxies, which are pervaded by 
far ultraviolet (FUV) radiation that destroys the molecules by photodissociation. 
In the diffuse interstellar medium (ISM) of galaxies like the Milky
Way, this radiation
is sufficient to maintain the gas occupying most of the volume
in an atomic or ionized state. Sufficient
concentrations of gas are able to exclude the dissociating radiation, both by
self-shielding and by dust absorption, and thereby become molecular;
these are the molecular clouds.
Regions in and around molecular clouds in which the thermal and chemical
state of the gas is determined by FUV radiation are
termed photodissociation regions (PDRs; e.g.\ \citealp{hol99}). 
The surface density of atomic gas in a PDR,
$\Sigma_\hone$, sets the threshold for gas to become
molecular and thus to be able to form stars.

The study of PDRs is a vast field, but the majority of the work that has been done is not suited to the problem of making general statements about the thicknesses of HI layers. Most of it is highly numerical  \citep{federman79, vandishoeck86, black87, dra96, neufeld96, sto96, 
spaans97, liszt00, liszt02, bro03}, which is useful for trying to match detailed properties of particular regions, but makes it difficult to draw general conclusions that allow extrapolation beyond the particular numerical parameters selected for a calculation.
The same problem of how to draw general conclusions from particular results arises with galactic-scale simulations that model H$_2$ formation and dissociation as one of many processes occurring in a galaxy \citep{robertson08, gnedin09}. In contrast, the analytic work preceeding this series of papers has been limited to the comparatively simple one-dimensional cases of a unidirectional beam of radiation impinging on a slab
 \citep{dra96, sternberg88} or a radially-converging radiation field striking the surface of a sphere \citep{elmegreen93};
the beam and slab case is a reasonable approximation for a PDR irradiated by a single star, but not for a cloud
immersed in the interstellar radiation field produced by the ensemble of many stars.
Since these authors assumed that the radiation is beamed,
they were able to introduce a shielding function that describes the
attenuation of the dissociating radiation with depth; this does not appear to be possible
for isotropic incident radiation, particularly for spherical clouds.
Moreover, these treatments did not identify the important dimensionless numbers that characterize the problem
of a spherical cloud immersed in an isotropic radiation field.
 Perhaps the paper that is closest in spirit to the present series of papers is that of
\citet{sternberg88}, who determined scaling laws for infrared fluorescent emission lines
of \htwo\
emitted by an irradiated slab.
He assumed that the radiation was beamed, so that
he could use a shielding function. 

We determined the dimensionless quantities that characterize irradiated
spherical clouds
 for the first time in \citet[hereafter Paper I]{kru08}, where we also gave an approximate solution for the atomic-to-molecular ratio of an irradiated gas cloud in terms of these numbers. In this paper we extend this analysis by introducing a new and more accurate method for solving the equations of radiative transfer and photodissociation. We also extend our treatment to the case of primordial H$_2$ formation, where there is no dust and H$_2$ instead forms via the H$^-$ channel. Finally, 
we apply our results to the detemination of the molecular fraction in atomic-molecular
complexes.

\subsection{The \strom\ Analysis}
PDRs are one example of a general class of problem in which one must find the chemical state of a gas cloud that simultaneously obeys coupled conditions of chemical and radiative equilibrium. The most famous example of this type of problem is that of determining the structure of an HII region, and the structure of PDRs and similar problems can be solved by adapting the classic solution to the HII region problem by \citet{str39}.
In the absence of dust,
the thickness of the layer of ionized gas created by an incident flux of
ionizing photons $F_{\rm ion}^*$ is
\beq
\ell_{\rm St}=\frac{F_{\rm ion}^*}{\alpha n_e^2},
\eeq
where the superscript $^*$ indicates photon quantities, $n_e$ is the density
of electrons, and $\alpha$ is the radiative recombination coefficient.
Similarly, for a point source of ionizing radiation, the radius of the
ionized gas (the \strom\ radius) is given by
\beq
R_{\rm St}=\left(\frac{3L_{\rm ion}^*}{4\pi \alpha n_e^2}\right)^{1/3}.
\eeq
Note that these results are independent of the value of the cross
section, $\sigma_{\rm ion}$, for absorption of the ionizing radiation, which is the process that actually
produces the ionized gas. 
However, in order for this simple \strom\ analysis to
be useful 
in determining $R_{\rm St}$, 
it is necessary for the thickness of the layer in which the
gas switches from primarily ionized to primarily atomic to be thin:
$R_{\rm St}\gg 1/n({\hone})\sigma_{\rm ion}$, which is generally satisfied for
HII regions around OB stars \citep{str39}. 
If this condition is not satisfied, the \strom\ analysis still provides the
total volume emission measure, $\int n_e^2 dV=L_{\rm ion}^*/\alpha$.
The effects of dust absorption
on \strom\ layers and spheres can  be readily included \citep{pet72}.

The same type of analysis can be applied to the atomic gas in a PDR
(Paper I).
This is particularly useful because, whereas photoionization
in an HII region is a relatively simple continuum absorption process,
photodissociation is a line absorption process; 
sophisticated calculations (e.g., \citealp{dra96}) are needed to
follow the absorption as the lines become
opaque and even overlap. Only a fraction $\fdiss$ of the line absorptions
result in dissociation of the \htwo\ molecule; the rest result in the emission
of fluorescence radiation. ($\fdiss$ is a weak function of frequency
and of the level populations of the H$_2$ molecules; the average value is in
the range 0.12---\citealp{dra96}---to 0.11---\citealp{bro03}, although it may be higher in regions within a PDR where a large fraction of the H$_2$ molecules are in excited states.)
The process that is analogous to radiative recombination
is molecule formation, which occurs at a rate $\fone\nh^2\calr$,
where $\nh$ is the number density of hydrogen nuclei,
$\fone\equiv n(\hone)/\nh$ is the \hone\ fraction, and $\calr$ is
the rate coefficient for \htwo\ formation. In the Galaxy, \htwo\ forms
on dust grains and the rate coefficient is $\calr\simeq 10^{-16.5}$~cm$^3$~s\e\
\citep{dra96}. In primordial, metal-free gas, \htwo\ forms via the
H$^-$ process
or the three-body process, 
which can be described similarly (\S 3).
In the absence of dust absorption, a point source with
a photon luminosity $L^*$ in the Lyman-Werner 
bands between 91.2~nm and 110.8~nm 
that is embedded in a medium of constant density, $\nh$,
can create a sphere of atomic gas
with a total number of neutral hydrogen atoms
\beq
\caln_\honet=\int f_1\nh dV=\frac{\fdiss L^*}{\calr\nh}.
\eeq
In the absence of dust, the transition between the atomic and molecular
gas is not sharp (Paper I); nonetheless, we can define a characteristic radius
of the atomic gas by setting $f_1=1$, obtaining
\beq
R_\hone=\left(\frac{3\fdiss L^*}{4\pi\nh^2\calr}\right)^{1/3},
\eeq
which has the same form as the \strom\ radius.
Similarly, a dissociating photon flux $F_\inc^*$ incident on the surface 
of a slab of gas can maintain an HI 
column density
\beq
N_\honet =\int f_1\nh dz=\frac{\fdiss F_\inc^*}{\nh\calr}.
\label{eq:nhonet}
\eeq
Setting $f_1=1$ gives the characteristic size of this layer,
\beq
\ell_{\hone,\, \rm slab}=\frac{\fdiss F_\inc^*}{\nh^2\calr}.
\eeq

An important limitation of the \strom\ analysis is that the bandwidth of the
dissociating radiation must be known in advance. As shown by \citet{sternberg88},
the effective bandwidth begins to shrink for metallicities $Z\ga Z_\odot$, where dust
absorption is sufficiently strong that the individual absorption lines no longer overlap.
For the cases of greatest interest, however, $Z\la Z_\odot$ and the results presented here
are applicable.

\subsection{Summary of Previous Papers in this Series}

In Paper I, we gave an approximate determination of the thickness of the HI layer
in a sphere irradiated from the outside by isotropic radiation.
For such a sphere, the incident flux is unknown, since the flux due to radiation
striking the surface from outside the sphere can be compensated in part by the
flux due to radiation that passes through the sphere. It is therefore advantageous
to characterize the radiation field by the ambient value of the
mean density of dissociating photons. 
If we let $E$ be the energy density of photodissociating radiation,
then $E^*$ is the corresponding photon density; let $E_0^*$ be the 
corresponding ambient value.
We define the characteristic value of the thickness of the HI layer as 
$\ell_{\hone,\, \rm slab}$ with $F_\inc^*$ replaced by $cE_0^*$:
\beq
\lch\equiv \frac{\fdiss cE_0^*}{\nh^2\calr}.
\label{eq:lch}
\eeq
For an opaque slab embedded in an isotropic radiation field, 
the energy density just outside the slab is half the ambient
value and the flux is half that, or $F_\inc^*=\frac 14 cE_0^*$;
as a result, $\ell_{\hone,\, \rm slab}\simeq \frac 14\lch$ in this case. The importance of dust
is measured by the characteristic dust optical depth,
\beq
\tau_{\rm \,ch}=\nh\sigma_d \lch=\frac{\fdiss\sigma_d cE_0^*}{\nh\calr}\equiv\chi,
\eeq
where $\chi$ is one of the two dimensionless parameters introduced in Paper I.
Photodissociation regions are often characterized by the ratio $G_0'/\nh$, where
$G_0'$ is the ratio of the dissociating radiation field, $E_0^*$, to the typical
value in the Milky Way, which is $7.5\times 10^{-4}$ photons cm\eee\ s\e\
from \citet{draine78}. In terms of this ratio, $\chi$ is
\beq
\chi=71\left(\frac{\sigma_{d,-21}}{\calr_{-16.5}}\right)\frac{G_0'}{\nh}.
\eeq
Physically, $\chi/f_1$ is the ratio of the number of LW photons absorbed by dust to the
number absorbed by \htwo\ molecules.
Thus, dust has a significant impact on the structure of the HI layer for $\chi\ga 1$.
The parameter $\chi$ is of order unity in the Milky Way (\citealp{kru09a}, hereafter Paper II;
see eq. \ref{eq:chigal} below). Clumps in the PDRs discussed by
\citet{hol99} have $\chi\sim 0.3-10$, while the interclump gas in those PDRs has
values about 100 times larger.
Observe that for the case in which \htwo\ forms on dust grains, both
$\sigma_d$ and $\calr$ are effectively measures of the total surface area of dust grains mixed into the gas; $\sigma_d$ measures the area available for absorbing photons, and $\calr$ the area available for adsorbing hydrogen atoms. Thus their ratio should be independent of metallicity, dust-to-gas ratio, density, and most other quantities.
For an opaque slab, we have seen that the incident flux $F^*_\inc=\frac 14 cE_0^*$,
so that $\ell_\hones=\frac 14\lch$ and
\beq
\tau_\hones=
\nh \sigma_d\ell_\hones=
\frac 14\chi\; ;
\label{eq:tauhones}
\eeq
keep in mind that these simple estimates are based on neglecting
absorption of the incident FUV radiation by dust. 
For future reference, we note that this implies that the total HI column density in a 
dust-free slab (eq. \ref{eq:nhonet}) is $N_\honet=\chi/(4\sigma_d)$.
Finally, we note that
the ratio of $\lch$ to the cloud radius, $R$, is
\beq
\frac{\lch}{R}=\frac{\fdiss cE_0^*}{\nh^2\calr R}=\frac{\chi}{\tau_R},
\eeq
where $\tau_R=\nh\sigma_d R$, the dust optical depth associated with the cloud
radius, is the other dimensionless parameter in Paper I.

As in the case of HII regions, a  \strom-type analysis is particularly useful
if the transition from atomic to molecular gas is sharp, $\ell_\hone\gg 1/(\nh\sigma_\mhtwo)$. 
In Paper I we showed that this transition is indeed relatively
sharp for $\chi\ga 1$, so that in that case it is possible to infer the size of the
atomic gas by setting $f_1\simeq 1$.
For the case with dust, we make this approximation here; this is the principal approximation in
our work. In the absence of dust, we can determine the total mass of \hone\ (in the spherical
case) or column of \hone\ (in the slab case) without making this approximation.

The advance made in Paper II was to recognize that the density in the
atomic gas in the transition region is set by the requirement that it
be in two-phase equilibrium \citep{wol03}. The latter authors showed that
the minimum density of cold HI (the CNM) in galaxies is
\beq
n_{\rm H,\, min}=\frac{31}{1+3.1Z^{\prime\,0.365}}\left(\frac{E_0^*}{7.5\times 10^{-4}
\;\mbox{cm\eee}}\right)~~~\mbox{cm\eee},
\eeq
where $E_0^*$ and $Z'$ are the radiation field and metallicity normalized to the local Milky Way value. 
If the density is several times the minimum value
($\nh=\phi_{\rm CNM} n_{\rm H,\, min}$ with $\phi_{\rm CNM}\sim 3$), 
then 
\beq
\label{eq:chieqn}
\chi=3.1\left[\frac{\sigma_{d,\,-21}}{\calr_{-16.5}(\phi_{\rm CNM}/3)}\right]
\left(\frac{1+3.1Z^{\prime\, 0.365}}{4.1}\right)
\label{eq:chigal}
\eeq
depends primarily on metallicity, and weakly at that; here
$\sigma_{d,\,-21}=\sigma_d/(10^{-21}~$cm$^2$), etc.
The surface density of \hone\ in a galaxy, assuming that it is in a slab and
neglecting dust absorption, is then
\beqa
\Sigma_\hone&=&\frac{\muh}{\sigma_d}\tau_\hones\nonumber,\\
	&\simeq&\frac{8.8}{Z'(\phi_{\rm CNM}/3)}\left(\frac{1+3.1Z^{\prime\, 0.365}}{4.1}\right)
	~~~M_\odot~\mbox{pc\ee}~~~
\eeqa
for $\calr_{-16.5}/\sigma_{d,\, -21}=1$; here $\muh=2.34\times 10^{-24}$~g is the mass
per hydrogen for a gas of cosmic abundances. 
Paper II showed that
dust absorption reduces this by about a factor of 2. On the other hand,
if the slab of gas is illuminated from both sides, then the total column
of \hone\ is increased by a factor 2. 
Paper II generalized these results to
the case of finite clouds and demonstrated that the results were in good agreement
with existing observations. Indeed, since the theory applies to individual cloud
complexes, predictions were made on how the results would change as 
the resolution of the observations improves.

\citet{kru09b} applied the results of Papers I and II together with
the earlier work of \citet{kru05} to determine the star formation
rate in galaxies as a function of metallicity. For low surface densities
($\Sigma\la 10/Z'~M_\odot$~pc\ee), the rate is dominated by the
transition from \hone\ to \htwo; for intermediate columns
($10/Z'~M_\odot~$~pc\ee$\;\la \Sigma\la 85~M_\odot$~pc\ee, it is determined
by the properties of giant molecular clouds; and for larger column densities
it is determined by the pressure of the galactic ISM.
Similarly, \citet{kru09c} showed how the results of Papers I and II can be used to explain the observed absence of damped Lyman-$\alpha$ systems with high column densities and metallicities.
The treatment of
the \hone---\htwo\ transition in both these papers relied on an analysis that is more broadly applicable
than the one developed in Paper I. We provide that analysis here.

	In \S 2, we present the general formalism for determining the
thickness of the \hone\ shielding layer for spherical molecular clouds.
This formalism is based on a \strom--type analysis. In \S 3, we present
the results for both slabs and spheres in the absence of dust; this is
relevant to the study of primordial clouds. \S 4 gives the results including
dust absorption,
\S 5 shows how these results can be generalized to the case of clouds in which the atomic and molecular regions have differing densities, and the conclusions are given in \S 6.

\section{Formalism}

\subsection{Basic Equations}

Consider a spherical cloud of radius $R$ and
uniform density exposed to an isotropic UV radiation field.
We describe the radiation field in terms of the specific intensity in photon units,
$\ins\equiv I_\nu/h\nu$. Let $\kappa_{2\nu}=n_2\stn$ be the opacity due to absorption
by \htwo\ molecules, where $n_2$ is the density of \htwo\ molecules and
$\stn$ is the \htwo\ absorption cross section at frequency $\nu$. Let
$\kappa_d=\nh\sigma_d$ be the opacity due to absorption by dust,
where $\nh$ is the density of H nuclei and $\sigma_d$ is the dust absorption cross
section per H in the photodissociation part of the spectrum, from 91.2 nm to
110.8 nm; we ignore the weak frequency dependence of this cross section.
The equation of transfer is then
\beq
\veceh\cdot\grad I_\nu^*=-(\kappa_{2\nu}+\kappa_d) I_\nu^*=-(n_2\stn+\nh\sigma_d)\ins
\eeq
for the intensity in the $\veceh$ direction.
The opacity $\kappa_{2\nu}$ depends
 on position, but we have suppressed the
argument for clarity.
For a cloud bathed in a uniform radiation field
of intensity $I_{\nu 0}^*$, the intensity at a point inside the cloud is 
\beq
\ins=I_{\nu 0}^*\exp(-\tau_{2\nu}-\tau),
\eeq 
where the optical depths are proportional to the column density:
the dust optical depth is
\beq
\tau=\int \nh\sigma_d ds
\eeq
and
the \htwo\ optical depth is
\beq
\tau_{2\nu}=\int n_2\stn ds = N_2\stn,
\eeq
where the range of integration extends along the ray from where the ray  enters 
the cloud to the point in question and $N_2$ is the corresponding
\htwo\ column density along the ray. 

We next integrate over the range of frequencies that can photodissociate \htwo, 
$\nu_2\geq\nu\geq\nu_1$, corresponding to wavelengths $91.2~$nm$\;\leq\lambda\leq 110.8$~nm.
The frequency-integrated equation of transfer is 
\beq
\veceh\cdot\grad I^*=-[\kappa_I(\veceh)+\kappa_d]I^*,
\eeq
where
\beq
\kappa_I(\veceh)=\frac{1}{I^*}\int_{\nu_1}^{\nu_2}\kappa_{2\nu}\ins d\nu.
\eeq
We include the argument $\veceh$ in $\kappa_I$ to emphasize that this
opacity depends on angle through its dependence on $\ins$;
$\kappa_I$ also depends on position.
The intensity-weighted \htwo\ opacity is
then a function of $N_2$,
\beq
\kappa_I=\frac{\int_{\nu_1}^{\nu_2}\kappa_{2\nu} \exp(-\tau_{2\nu})\, d\nu}
{\int_{\nu_1}^{\nu_2}\exp(-\tau_{2\nu})\, d\nu}=\kappa_I(N_2).
\label{eq:stin}
\eeq
The dependence of $\kappa_I$ on $N_2$ is governed by two countervailing factors. On one hand, as $N_2$ increases more and more photons at frequencies near line centers are absorbed. Since $\stn$ decreases away from line center, the mean the opacity per H$_2$ molecule is reduced. On the other hand, $\kappa_{2\nu}$ is also proportional to the density of H$_2$ molecules 
$n_2$, 
which
increases with $N_2$.
The intensity-weighted optical
depth that determines the attenuation of $I^*$ is $\tau_I=\int \kappa_I \, ds$,
so that
\beq
I^*=I_0^*\exp[-\tau_I+\tau].
\label{eq:is}
\eeq

	In terms of the first three angular moments of the radiation field,
\beq
E_\nu^*\equiv \int d\Omega\; \ins,~~~\vecF_\nu^*\equiv \int d\Omega\; \veceh\ins,~~~
\rvecP=\frac{1}{c}\int d\Omega\;\veceh\veceh\ins,
\eeq
the first two angular moments of the equation of radiative transfer are
\beqa
\div\vecF_\nu^*&=&-c(\kappa_{2\nu}+\kappa_d) E_\nu^*,\\
\div\rvecP_\nu^*&=&-\frac{1}{c}(\kappa_{2\nu}+\kappa_d) \vecF_\nu^*.
\eeqa
Integrating over frequency gives
\beqa
\div\vecF^*&=&-c(\kappa_E +\kappa_d)E^*,
\label{eq:divf}\\
\div\rvecP^*&=&-\frac{1}{c}(\kappa_F+\kappa_d) \vecF^*,
\label{eq:divp}
\eeqa
where
\begin{eqnarray}
\kappa_E & =& \frac{1}{E^*}\int_{\nu_1}^{\nu_2} d\nu \kappa_{2\nu} E_\nu^* \nonumber \\
& =& \frac{1}{E^*}
\int d\Omega\int_{\nu_1}^{\nu_2} d\nu \kappa_{2\nu} \ins=\frac{1}{E^*}\int d\Omega\; \kappa_I I^*.
\label{eq:kappaedef}
\end{eqnarray}
We have assumed that the direction of $F_\nu^*$ is independent of frequency;
as a result, $\kappa_F$ is given by an expression like that for $\kappa_E$ with
$E^*$ replaced by $|\vecF^*|$ and with a factor $\mu$ in the angular integrals. 
Writing out equations (\ref{eq:divf}) and (\ref{eq:divp}) in
the case of spherical symmetry, we obtain
\beqa
\frac{1}{r^2}\frac{d}{dr}\, r^2F^* &=&c(\kappa_E+\kappa_d) E^*,
\label{eq:divfs}\\
\frac{dP^*}{dr}+\frac{3P^*-E^*}{r}&=&\frac{1}{c}(\kappa_F +\kappa_d)F^*,
\label{eq:divps}
\eeqa
where $F^*\equiv |\vecF^*|=-F_r^*$ and $P^*= P^*_{rr}$ is the $rr$ component
of the radiation pressure tensor.

	Now define the characteristic flux
\beq
\fcs\equiv \frac{\nh^2\calr R}{\fdiss},
\eeq
which is the flux required to photodissociate a layer of thickness $R$, as
discussed in \S 1.
The solution is determined by the two quantities defined in Paper I,
\beqa
\chi&\equiv&\frac{\fdiss \sigma_d c E_0^*}{\nh\calr},\\
\tau_R&\equiv& \nh\sigma_d R,
\eeqa
where $E_0^*$ is
the number density of dissociating photons far from the cloud.
In terms of these parameters, we have
\beq
\fcs=\left(\frac{\tau_R}{\chi}\right) cE_0^*.
\label{eq:fcs}
\eeq
We define the normalized flux and photon density by
\beq
\fsh\equiv \frac{F^*}{\fcs},~~~~~\esh\equiv\frac{cE^*}{\fcs}.
\label{eq:esh}
\eeq
The normalized photon density far from the cloud is then
\beq
\eosh=\frac{\chi}{\tau_R}.
\label{eq:eosh}
\eeq

	We now use the assumption that the formation and destruction of
\htwo\ are in balance in order to determine the opacity $\kappa_E$;
this is the essence of the \strom\ argument.
Observe that it is difficult to calculate $\kappa_E$ explicitly, since it involves
first a frequency integration over the cross section and the specific intensity, both of which are
complicated functions of frequency due to line absorption, and then an average
over angle.
Balancing the rates of formation and destruction of \htwo\ gives
\beqa
f_1\nh^2\calr&=&\fdiss\int d\Omega\int_{\nu_1}^{\nu_2} d\nu \;\kappa_{2\nu}\ins,\\
&=&\fdiss c\kappa_E E^*,
\label{eq:balance}
\eeqa
where $f_1\equiv n(\mbox{HI})/\nh$ is the 
atomic fraction and $\calr$ is the rate coefficient for \htwo\ formation. 
The principal approximation in our analysis is to assume that the transition
from atomic to molecular gas is sharp, so that $f_1\simeq 1$ in the atomic gas.
The balance equation then implies that
\beq
\kappa_E=\frac{\nh^2\calr}{c\fdiss E^*}=\frac{1}{R\esh}.
\label{eq:kappa}
\eeq
Our seond approximation is to relate the flux-weighted opacity to $\kappa_E$ by
\beq
\kappa_F=\phik\kappa_E,
\eeq
where $\phik$ is a constant. The results for the 1D case below verify that this
is a good approximation. We expect $\phik$ to exceed unity since the rays that
carry the flux are closer to the radial direction and have therefore undergone less
attenuation and interact with a larger cross section, as mentioned below equation
(\ref{eq:stin}).

To close the moment equations of radiative transfer, 
we introduce a variable Eddington factor,
\beq
\fedd\equiv \frac{\hat P^*}{\esh}.
\eeq
Expressing equations (\ref{eq:divfs}) and (\ref{eq:divps}) in
dimensionless form and using equation (\ref{eq:kappa}) to eliminate $\kappa_E$, we obtain
\beqa
\frac{d\fsh}{dx}+\frac{2\fsh}{x}&=& f_1
+\tau_R\esh,
\label{eq:fsh}\\
\frac{d}{dx}\,(\fedd\esh)+\frac{(3\fedd-1)\esh}{x}&=&\frac{\phik f_1
\fsh}{\esh}
+\tau_R\fsh,
\label{eq:feddesh}
\eeqa
where $x\equiv r/R$. These equations must be solved subject to the boundary conditions that 
both the photon density and the flux vanish at $x_{{\rm H_2}}$, the transition point
from atomic to molecular gas; i.e., $\esh(x_{{\rm H_2}})=\fsh(x_{{\rm H_2}})=0$.
The values of the photon density and flux at the surface are not known,
except in the case in which the cloud is completely opaque.

\subsection{Approximate Determination of the Eddington Factor $\fedd$}
\label{sec:edd}

	Many astrophysical problems are characterized by a central source of
radiation in an opaque region, where the radiation is nearly isotropic, surrounded
by a less opaque region in which the radiation is beamed outwards. Shu (1991)
writes a general form for a relation to close the moment equations of radiative 
transfer as
\beq
E^*=3P^*-\frac{q_F}{c}F^*.
\label{eq:shu}
\eeq
He points out that the choice $q_F=2$ makes a smooth transition from
the isotropic case ($E^*=3P^*,~F^*\rightarrow 0$) to the beamed case
($E^*=P^*=F^*/c$). Our problem is quite different, however: the radiation
is isotropic at large distances from the cloud, but close to the cloud there
is a ``hole" in the radiation field since radiation cannot pass through the molecular
core. We therefore need to develop a different approach to closure.

\begin{figure}
\plotone{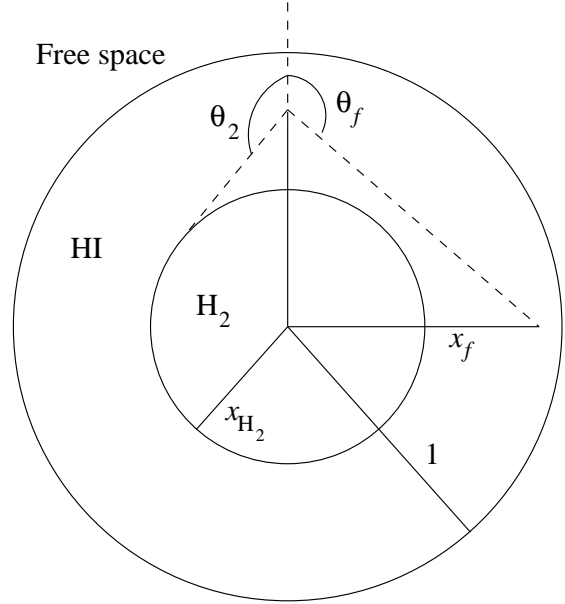}
\caption{
\label{diagram}
Diagram of the geometry of a spherical cloud illuminated by an external radiation field.
}
\end{figure}

	As remarked above, we close the moment equations by using 
variable Eddington factor, $\fedd$. Our approximation for $\fedd$ 
at any point inside the cloud is based on 
representing the main qualititive feature of the radiation field---that 
the intensity decreases for rays at larger angles to the outward normal
and vanishes for rays that have intersected the molecular core---by 
a step function decrease at some angle $\theta_f$. 
(We adopt
the convention in Paper I that $\theta$ is the angle between
the outward normal and the direction from which the photon originates
 -- see Figure \ref{diagram}.)
We thus approximate
the radiation field as being of constant intensity for $0\leq\theta<\theta_f$ 
(corresponding to $1\geq\mu\geq \mu_f$, where $\mu\equiv\cos\theta$) and
vanishing for $\theta_f<\theta\leq \pi$ (corresponding to $\mu_f>\mu\geq -1$). We then obtain
\beq
\fedd\equiv\frac{P^*}{E^*}=\frac{\int_{\mu_f}^1 \mu^2 d\mu}{\int_{\mu_f}^1 d\mu}
=\frac 13(1+\mu_f+\mu_f^2).
\label{eq:feddo}
\eeq

It remains to adopt an {\it ansatz} for $\mu_f$. 
First consider the dust-free case.
For points close
to the molecular core $(x\rightarrow x_{{\rm H_2}})$, we assume that
$2\theta_f$ is the angle subtended by the molecular core, so
that $\sin\theta_f\simeq x_{{\rm H_2}}/x$. In the opposite limit, in which
the molecular core is assumed to be small ($x_{{\rm H_2}}\ll 1$) and the
point is far from the core, we assume that a ray at $\theta_f$ intersects
the midplane at a distance $x_f R$ from the center of the cloud,
so that $\tan\theta_f\simeq x_f/x$. We combine these two limits by
assuming that $\tan^2\theta_f$ is the sum of the squares of the
tangents in the two limiting cases,
\beq
\tan^2\theta_f \equiv \frac{x_{{\rm H_2}}^2}{x^2-x_{{\rm H_2}}^2}+\frac{x_f^2}{x^2}.
\eeq
The corresponding cosine is $\mu_f=-1/(1+\tan^2\theta_f)^{1/2}$,
where the minus sign reflects the fact that the angle is in the second
quadrant:
\beq
\mu_f(x)=-\left[\frac{x^2(x^2-x_{{\rm H_2}}^2)}{x^4+x_f^2(x^2-x_{{\rm H_2}}^2)}\right]^{1/2}.
\label{eq:muf}
\eeq
The parameter $x_f$ is an eigenvalue of the problem,
which is adjusted to give the correct net flux at the surface.

	How does this compare with Shu's (1991) closure approximation?
Using the model that the radiation field is uniform for $\mu>\mu_f$ and
zero for $\mu<\mu_f$ implies that the flux is $\frac 14 E^*(1-\mu_f^2)$.
Solving for the closure factor $q_F$ in equation (\ref{eq:shu}), we find
$q_F=2\mu_f$. Since $\mu_f$ is negative, it follows that the
closure factor is negative, which is qualitatively different than
the standard case. The fact that $\mu_f<0$ implies that
$\fedd$ in equation (\ref{eq:feddo}) is always {\it less} than $\frac 13$;
it is in the range $\frac 13\geq\fedd\geq\frac 14$.

	Next, consider the case in which dust is included. For a dusty slab,
the dissociating photon density at an optical depth $\tau$ from the surface is
\beq
E^*=\frac 12 E_0^*
\int_0^1\exp(-\tau/\mu)d\mu,~~~~~\mbox{(dusty slab).}
\eeq
Evaluation of the integral
gives
\beq
E^*=\frac 12 E_0^*E_2(\tau)\simeq\frac 12 E_0^*\;\frac{e^{-\tau}}{1+\tau},
\eeq
where $E_2$ is an exponential integral and where the approximation is accurate
to about 25\%. The advantage of the approximation is that it shows that
the photon density is as if the intensity along the normal filled a solid
angle corresponding to $\Delta\Omega=2\pi\Delta\mu=2\pi(1-\mu_d)$
with $1-\mu_d=1/(1+\tau)$, so that
\beq
\mu_d=\frac{\tau}{1+\tau}.
\eeq
For the spherical case, we take $\tau = (1-x)\tau_R$, where $\tau_R$ is the dust optical depth from the cloud center to edge.
To join the case where dust dominates the opacity onto the case where molecular absorption dominates it, we use the simple {\it ansatz}
\beq
\fedd =\frac 13(1+\mu_t+\mu_t^2).
\label{eq:fedd}
\eeq
with $\mu_t = \mu_f + \mu_d$.
Note that whereas $\mu_f$ is always negative in the dust-free case, it
can be positive when dust is important.

\subsection{Boundary Conditions for the Spherical Case}

\label{sec:sphere}

	As in Paper I, we use the solution of the radiative transfer equation at the
cloud surface to constrain the solution. 
At the surface ($x=1$), photons at $\mu>0$ are unattenuated, those that
traverse the cloud outside the molecular core, $0>\mu>\mu_2$,
with
\beq
\mu_2=-\left(1-x_{{\rm H_2}}^2\right)^{1/2},
\eeq
are attenuated, and those that strike the molecular core ($\mu<\mu_2$) are absorbed.
Integration of equation (\ref{eq:is}) then gives
\beq
\esh(\tau_R)=\frac 12 \eosh \left[1+\int_{\mu_2}^0 d\mu\; e^{-(\tau_I+\tau)}\right].
\eeq
We then introduce our next approximation: we replace $\tau_I=
\int \kappa_I(r,\mu) ds$, with
$\tau_E=\int\kappa_E(r) ds$
(see eq.\ \ref{eq:kappaedef}). Note that the error introduced by this approximation is
limited, since the term in which it is made contributes at most half the total.
Proceeding in the same fashion with the evaluation
of the flux at the surface, we obtain
\beq
\frac{\esh(\tau_R)}{\fsh(\tau_R)}=
2\left(\frac{1+\int_{\mu_2}^0 d\mu\; e^{-(\tau_E+\tau)}}{1+2\int_{\mu_2}^0 d\mu\; \mu
e^{-(\tau_E+\tau)}}\right),
\label{eq:constraint}
\eeq
where the factor 2 in front arises since the unattenuated intensity is $\frac 12 E_0^*$
whereas the unattenuated flux is $\frac 14 cE_0^*$.
The optical depth is given by
\beq
\tau_E=2\int_{\sin\theta}^1\frac{x'dx'}{(x'^2-\sin^2\theta)^{1/2}\esh},
\eeq
where $\sin\theta$ is the value of $x$ of closest approach to the center 
of the cloud for a ray at an angle $\theta$.

The equations we have now written down constitute a complete set that fully determines the solution. For a given choice of $\chi$ and $\tau_R$, a solution consists of two unknown numbers, $\mu_f$ (or equivalently $f_{\rm Edd}$, since the two are related by equation \ref{eq:fedd}) and $x_{\rm H_2}$, and two unknown functions, $\esh$ and $\fsh$. The two functions are constrained by the ordinary differential equations (\ref{eq:fsh}) and (\ref{eq:feddesh}), while the two numbers are constrained by the algebraic equations $\esh(x_{\rm H_2})=\fsh(x_{\rm H_2})=0$ and the consistency condition equation (\ref{eq:constraint}). We must choose a fixed value of $\phik$, since we lack an equation to determine an additional parameter. We discuss how to choose this value in \S~\ref{sec:slab}, and we defer discussion of how to obtain the solution numerically once we have chosen $\phik$ until \S~\ref{sec:prisphere} and \S~\ref{sec:dustysphere}.

\subsection{The Semi-Infinite Slab Limit}

\label{sec:slab}

	The normalization we have used for the spherical case, which is
based on $\fcs\propto R$, breaks down for a one-dimensional,
semi-infinite slab, which corresponds
to the limit $R\rightarrow\infty$. We therefore normalize with respect to
the ambient radiation field, $E_0^*$ by defining
\beq
\fsho\equiv\frac{F^*}{cE_0^*},~~~~~\esho\equiv\frac{E^*}{E_0^*};
\eeq
thus, $\fsho(0)=\frac 14$ and $\esho(0)=\frac 12$ at
the surface of the slab.
The first two moments of the radiative transfer equation become:
\beqa
\frac{d\fsho}{d\tau}&=&-\frac{f_1}{\chi}-\esho,
\label{eq:dfsh}
\\
\frac{d(\fedd\esho)}{d\tau}&=&-\frac{\phi_\kappa 
f_1
\fsho}{\chi\esho}-\fsho,
\label{eq:dfesh}
\eeqa
where $d\tau=\nh\sigma_d dz$, with $z$ being zero at the surface and increasing inward.
In the 1D limit ($x,\,x_{{\rm H_2}}\rightarrow 1$), the angle $\mu_f$ that enters the Eddington 
factor approaches zero (eq. \ref{eq:muf}). The Eddington factor thus becomes
\beq
\fedd=\frac 13 (1+\mu_d+\mu_d^2).
\label{eq:feddo1}
\eeq
For small $\tau$ (i.e., close to the surface), $\fedd\rightarrow \frac 13$
as it should, since the radiation is approximately isotropic in the half-space $\mu>0$.
For large $\tau$, we have $\mu_d=\tau/(1+\tau)\rightarrow 1$, and $\fedd\rightarrow 1$; this too is as it should be, since highly extincted radiation is beamed so that $P^*\simeq E^*$.

As with the spherical case, we have now written down enough equations to close the system. The independent parameter that determines the solution is the dimensionless radiation intensity $\chi$. For a given $\chi$, a solution consists of two numbers, the ratio $\phik$ of the flux-mean opacity to the energy-mean opacity and the dust optical depth $\tau_{\rm H_2}$ at which the molecular transition occurs, plus two functions, $\esho$ and $\fsho$. These quantities are constrained by two algebraic equations, $\esho(\tau_{\rm H_2}) = \fsho(\tau_{\rm H_2}) = 0$, and the two ODEs (\ref{eq:dfsh}) and (\ref{eq:dfesh}). These ODEs are subject to the boundary conditions $\esho(0)=1/2$ and $\fsho(0)=1/4$. This set of constraints therefore fully specifies the system. We defer discussion of our solution algorithm to \S~\ref{sec:prislab} and \ref{sec:dustyslab}. We pause here to point out an important distinction between the spherical and slab cases: for the spherical case, the angle $\theta_f$ at which we transition from zero flux to finite flux represents an extra parameter to be determined that is not present in the slab case. As a result, for the slab case we can use our constraint equations to determine $\phik$, and we can then use the resulting value of $\phik$ in the spherical case. As we shall see, a single value of $\phik$ works quite well over a very broad range of radiation fields.

\section{The Dust-Free Case: Primordial Clouds}

Although the motivation for this series of papers was to understand the atomic-molecular
transition in galaxies today, where \htwo\ is formed on grains,
our methodology applies to primordial gas clouds also, where \htwo\ is formed
via gas phase reactions. The effective values of $\calr$ for gas-phase production
of \htwo\ are given in the Appendix. The results obtained here are also useful
for comparison with the results for dusty clouds in the following section.

\subsection{Semi-Infinite Slab}
\label{sec:prislab}
 
 In the limit that $\kappa_d\rightarrow 0$, equations (\ref{eq:dfsh}) and
 (\ref{eq:dfesh}) for the 1D case become
 \beqa
 \frac{d\fsho}{d\hat N_\hone}&=&-\frac 14,\\
 \frac{d\fedd\esho}{d\hat N_\hone}&=&-\frac{\phik\fsho}{4\esho},
\eeqa
where we used the fact that $\kappa_d/\chi=\nh\sigma_d/\chi=\nh/(4N_\honet)$ 
(see below eq. \ref{eq:tauhones}) and
defined
\beq
\hat N_\hone\equiv\frac{N_\hone}{N_\honet},
\eeq
which varies from 0 at the surface of the slab to 1 at the point that the slab becomes fully
molecular.
Note that we have not assumed anything about the spatial variation of the HI fraction, $f_1$.
As pointed out in \S\ref{sec:slab}, the boundary conditions are $\fsho(1)=\frac 14$
and $\esho=\frac 12$; we also pointed out that for
small $\tau$, the Eddington factor approaches $\fedd=\frac 13$.
In the dust-free case, the use of $\fedd=\frac 13$ at large optical depths in the
\htwo\ lines is justified by the fact that the photons that dominate the photodissociation
are in the line wings; they are not highly extincted and therefore are not beamed.
Integration of the equations then gives
\beqa
\fsho&=&\frac 14(1 -\hat N_\hone),\\
E_1^{*\,2}&=&\frac 14-\frac{3\phik}{8}\left(\hat N_\hone-\frac 12\hat N_\hone^2\right).
\eeqa
Since $\esho$ must vanish at the same point that $\fsho$ does
(i.e., at 
$\hat N_\hone=1$), 
it follows that
\beq
\phik=\frac 43.
\eeq
We shall see in \S\ref{sec:dustyslab} that this remains a good approximation
in the dusty case for all physically relevant values of $\tau$.

\subsection{Spherical Clouds}
\label{sec:prisphere}

The problem of the dust-free spherical cloud does not have an exact solution
with the \strom\ method because the net incident flux, $F^*_\inc$, depends on
the structure of the molecular hydrogen in the transition zone; some of the
rays that strike the cloud will penetrate all the way through, reducing the
incident flux by an amount that depends on the unknown spatial distribution of
the \htwo. We therefore consider two complementary approximations:
In the first method, we assume that the gas is entirely atomic in the transition
zone ($f_1=1$) and calculate the incident flux self-consistently. In the second method,
we allow for the fact that molecular gas exists in the transition zone ($f_1<1$),
but assume that the transition zone is thin enough that the incident flux 
has the value appropriate for an opaque cloud, $F^*_\inc=\frac 14 cE_0^*$. We shall see
that the results overlap, giving a self-consistent determination of the region of
validity of our results.

In general,
 both $\chi$ and $\tau_R$ go to zero for dust-free spherical clouds, since both are proportional to the dust cross section $\sigma_d$. However, the ratio
\begin{equation}
\xi \equiv \frac{\chi}{\tau_R} =\frac{\lch}{R}= \frac{f_{\rm diss} c E_0^*}{n_{\rm H}^2 \calr R}
\end{equation}
remains finite, and this becomes the independent parameter describing the normalized radiation intensity in the dust-free case. 
We anticipate that the cloud will be fully atomic at high values of $\xi$, whereas the
atomic gas will be confined to a thin shell at low $\xi$.

\subsubsection{Method 1: $f_1=1$}

With $\tau_R=0$
and $f_1=1$,
equation (\ref{eq:fsh}) reduces to
\begin{equation}
\frac{d\fsh}{dx} + \frac{2 \fsh}{x} = 1
\end{equation}
which may be integrated analytically, 
subject to the boundary condition $\fsh(x_{\rm H_2})=0$, 
to obtain
\begin{equation}
\fsh = \frac{x}{3}\left[1 - \left(\frac{x_{\rm H_2}}{x}\right)^3\right].
\end{equation}
Substituting this into equation (\ref{eq:feddesh}) gives
\begin{equation}
\label{eq:feddesh_nodust}
\frac{d}{dx} (f_{\rm Edd} \esh) + \frac{(3 f_{\rm Edd} - 1) \esh}{x} = \frac{\phik x \left[1-(x_{\rm H_2}/x)^3\right]}{\esh}.
\end{equation}

\begin{figure}
\plotone{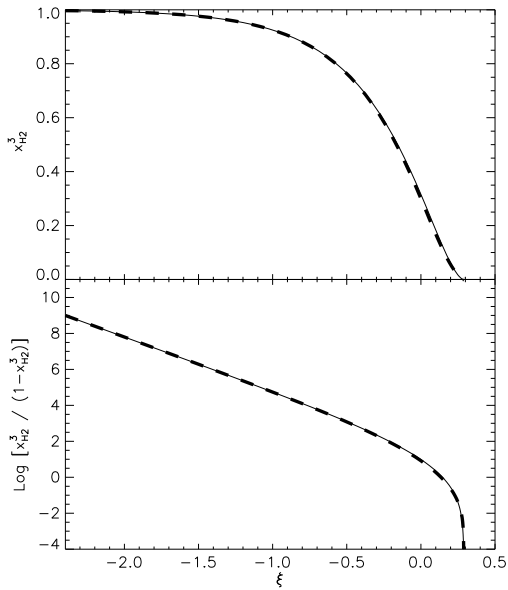}
\caption{
\label{fig:nodust_sphere}
Dust-free sphere.
Fraction of volume occupied by molecular gas $x^3_{\rm H_2}$ ({\it upper panel}) and ratio of molecular to atomic volume $[x_{\rm H_2}/(1-x_{\rm H_2})]^3$ ({\it lower panel}) as a function of $\xi\equiv\chi/\tau_R$. The solid line is the numerical solution, and the thick dashed line is the analytic approximation given by equation (\ref{eq:nodust_analyt}).
}
\end{figure}

We can now obtain the relationship between $x_{\rm H_2}$ and $\xi$ via the following algorithm:
\begin{enumerate}
\item  Select a value for $x_{\rm H_2}$. Our algorithm will determine the corresponding $\xi$.
\item Pick a trial value of the parameter $x_f$. (The value of $\phik$ is fixed to $4/3$, as discussed in \S~\ref{sec:prislab}.) The choice of $x_f$ determines the Eddington factor $f_{\rm Edd}$ via equation (\ref{eq:feddo}). 
\item Given $f_{\rm Edd}$, determine the function $\esh$ by integrating the ordinary differential equation (\ref{eq:feddesh_nodust}) from $x=x_{\rm H_2}$ to $x=1$, using the boundary condition that $\esh(x_{\rm H_2})=0$.
\item Check whether the solution for $\esh$ obeys the integral constraint equation (\ref{eq:constraint}) to within some specified tolerance. If not, repeat steps 2 -- 4, using Newton-Raphson iteration to find a value of $x_f$ that minimizes the difference between the two sides of equation (\ref{eq:constraint}).
\item If the solution does satisfy the constraint equation (\ref{eq:constraint}), equation (\ref{eq:eosh}) tells us that $\xi=\esh(1)$.
\end{enumerate}
Applying this algorithm yields the curve for $x_{\rm H_2}$ versus $\xi$ shown in Figure \ref{fig:nodust_sphere}. Note that 
in this approximation,
$x_{\rm H_2}$ goes to zero exactly for $\xi=2$.
In an exact calculation with $f_1<1$, the gas would not be fully atomic at this point,
but this result shows that clouds with $\xi>2$ are substantially atomic.
The formula
\begin{equation}
\label{eq:nodust_analyt}
x_{\rm H_2}^3 \simeq 1 - \left(\frac{3}{4}\right) \frac{\xi}{1 + 0.0712\,\xi^{2.8}}
\label{eq:x2dfsphere}
\end{equation}
matches the numerical solution to within 11\% in the range $x_{\rm H_2}^3>0.01$, and may be used in place of a numerical evaluation for most practical calculations. 

\subsubsection{Method 2: $F^*_\inc=\frac 14 c E_0^*$}
\label{sec:methodtwo}

If we retain the dependence on $f_1$ but assume that the incident flux
is known, an analytic solution is possible. 
In this case, equation (\ref{eq:fsh}) reduces to
\begin{equation}
\frac{d\fsh}{dx} + \frac{2 \fsh}{x} = f_1.
\label{eq:fsh1}
\end{equation}
The boundary condition $F^*_\inc=F^*(1)=\frac 14 cE_0^*$ becomes
\beq
\fsh(1)=\frac 14\xi
\eeq
with the aid of equation (\ref{eq:eosh}). Integration of equation (\ref{eq:fsh1})
gives
\beq
3\int_x^1 x^2f_1 dx=\frac{\caln_\hone}{\caln_{\rm H,\, cloud}}=3\left(\frac{\xi}{4}-x^2\fsh\right),
\label{eq:intx2}
\eeq
where $\caln_\hone$ is the number of HI atoms outside $r=xR$ and
$\caln_{\rm H,\, cloud}=4\pi R^3 \nh/3$ is the total number of H nuclei in the cloud.
To relate this result to that from the first method, we define an effective value
of $x_{\rm H_2}$ such that
\beq
x_{\rm H_2,\, eff}^3=1-\frac{\caln_\honet}{\caln_{\rm H,\, cloud}},
\eeq
where $\caln_\honet$, the total number of HI atoms, is evaluated at 
$x^2\fsh/(\xi/4)\rightarrow 0$. Equation (\ref{eq:intx2}) then implies
\beq
x_{\rm H_2,\, eff}^3=1-\frac 34 \xi.
\eeq
This has the same form as equation (\ref{eq:x2dfsphere}), which is
based on the first method, for
$\xi\la 1$. Since the first method properly accounts for the flux that penetrates
through the cloud, we conclude that the total number of HI atoms in a spherical,
dust-free cloud is 
\beq
\caln_{\honet}=\frac 34 \xi \caln_{\rm H,\, cloud}
\eeq
for $\xi\la 1$. 
Since Method 1 showed that the cloud becomes fully atomic for $\xi\ga 2$
and both methods agree for $\xi\la 1$,
we conclude that the results of this section 
are valid everywhere except where the cloud is close
to being fully atomic.

\section{Dusty Clouds 
($f_1=1$)}
\label{sec:dusty}

\subsection{Semi-Infinite Slab}
\label{sec:dustyslab}

\begin{figure}
\plotone{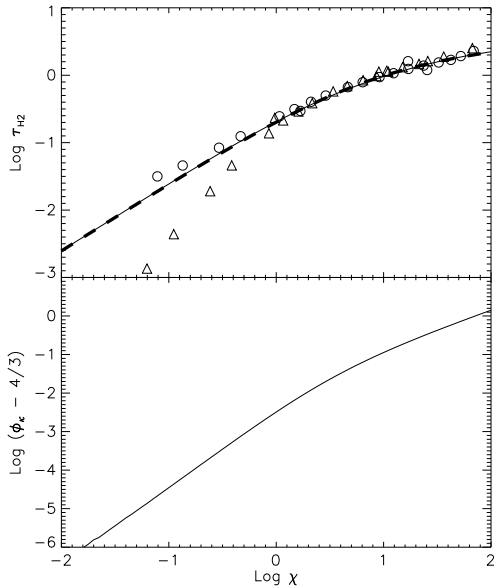}
\caption{
\label{fig:dustyslab}
{\it Upper panel:} dust optical depth of the atomic-to-molecular transition $\tau_{\rm H_2}$ as a function of radiation field $\chi$. The solid line is the numerical solution and the thick dashed line is the analytic approximation given by equation (\ref{eq:dustyslab_analyt}). Symbols show $\tau_{\rm H_2}$ versus $\chi$ for the Meudon runs, with circles showing the results if we adopt the integral definition of $\tau_{\rm H_2}$ and $\chi$, and triangles for the point definition. See \S~\ref{sec:slabcomparison} for details. {\it Lower panel:} $\log (\phik-4/3)$ as a function of $\chi$.
Paper II suggested that galaxies typically have $\chi\sim 2-3$ for metallicities
$Z\sim(0.1-1)Z_\odot$.
}
\end{figure}

We now consider the case of clouds with dust, $\sigma_d > 0$, beginning with the semi-infinite slab. 
As discussed in \S 1, the presence of dust sharpens the transition between atomic
and molecular gas, enabling us to make the approximation that $f_1=1$ in the atomic
gas.
Our goal is to determine the relationship between the imposed dimensionless radiation field $\chi$ and the dust optical depth $\tau_{\rm H_2}$ at which the transition from atomic to molecular occurs. In the process we must determine the unknown parameter $\phik$. We do so using an algorithm slightly modified from that in \S~\ref{sec:prisphere}:
\begin{enumerate}
\item Select a value for $\chi$. Our algorithm will determine the corresponding $\tau_{\rm H_2}$.
\item Pick a trial value of the parameter $\phik$. 
\item Given $\phik$, integrate the ordinary differential equations (\ref{eq:dfsh}) and (\ref{eq:dfesh}) using the boundary conditions that $\esho(0)=1/2$ and $\fsho(0)=1/4$. Integrate in the direction of increasing $\tau$, stopping once we reach a value $\tau=\tau_0$ for which $\fsho(\tau_0)=0$ or $\esho(\tau_0)=0$.
\item If $\fsho(\tau_0)=0$, check whether $\esho(\tau_0)=0$ as well to within some specified tolerance; if $\esho(\tau_0)=0$, check whether $\fsho(\tau_0)=0$ to within the specified tolerance. If not, repeat steps 2--4 using Newton-Raphson iteration on $\phik$ to minimize the value of $\esho(\tau_0)$ or $\fsho(\tau_0)$.
\item If $\esho(\tau_0)=\fsho(\tau_0)=0$ to within the specified tolerance, then $\tau_{\rm H_2} =\tau_0$.
\end{enumerate}
Applying this algorithm produces the curve for $\tau_{\rm H_2}$ as a function of $\chi$ shown in Figure \ref{fig:dustyslab}. The formula
\begin{equation}
\label{eq:dustyslab_analyt}
\tau_{\rm H_2} \simeq \ln \left[1 + \frac{\chi}{4(1+\epsilon)}\right],
\label{eq:tauh2}
\end{equation}
with $\epsilon=0.122\, \chi^{0.62}$, reproduces the numerical solution to within 2\% for $\log\chi<2.5$.

It is interesting to compare our result with that of \citet{sternberg88}, who wrote the
chemical balance equation (\ref{eq:balance}) as
\beq
f_1\nh^2\calr=D\fshield(N_2) n_2 e^{-(N_1+2N_2)\sigma_d},
\eeq
where $Dn_2(0)=\fdiss c\kappa_E(0)E^*(0)$ is the dissociation rate at the cloud surface
(this is a slight change from \citealp{sternberg88}, who defined $D$ as the
rate for the \citealp{draine78} radiation field). Integration of this equation,
without the assumption that $f_1\simeq 1$, gives
\beq
\sigma_d\int_0^\infty n_1 dz =\ln\left(1+\frac{DG}{\nh\calr}\right),
\label{eq:stern1}
\eeq
where $n_1$ is the atomic hydrogen number density and
\beq
G\equiv \sigma_d\int_0^\infty \fshield(N_2) e^{-2N_2\sigma_d} dN_2.
\eeq
Provided the transition from atomic to molecular gas is sharp, 
which is a good approximation in the dusty case
when $\chi \ga 1$ (Paper I),
then the LHS of equation (\ref{eq:stern1}) is just $\tau_\mhtwo$.
Sternberg's result then has the same form as ours, although it requires
knowledge of the shielding function $\fshield$
in order to evaluate it accurately. 
Furthermore, 
Sternberg considered the case of beamed radiation,
whereas in our problem the ambient radiation is isotropic far from
the cloud. 
For beamed radiation, $F^*=cE^*$, so that equation (\ref{eq:tauh2}) becomes
$\tau_{\rm H_2}\simeq \ln(1+\chi)$ (neglecting $\epsilon$); 
Sternberg's factor $DG/\nh\calr$ is thus
comparable to our factor $\chi$, and it is possible to show this directly
(Sternberg, private communication).
In Paper I we compared the results from our \strom\ analysis 
for beamed radiation with those from a detailed numerical calculation of
\htwo\ photodissociation and found good agreement. Our results are therefore
consistent with those of Sternberg.

It is also interesting to examine the value of $\phik$ determined by our procedure. For the dust-free case  $\phik=4/3$ exactly, and the lower panel of Figure \ref{fig:dustyslab} shows that this remains a very reasonable approximation even when $\chi$ is significant, meaning that dust opacity plays an important role in shielding molecules. The approximation $\phik=4/3$ is good to better than 10\% for $\chi<12.6$, and good to within 50\% for $\chi < 50$. Since typical molecular clouds have $\chi\simeq 1$ (Paper II), this means that we can adopt $\phik=4/3$ in general without significant error for the parameters that describe real clouds.

\subsubsection{Comparison to Numerical Results}
\label{sec:slabcomparison}

The dusty slab case is a useful one to consider because there are a number of PDR codes that solve for the transition for \hone\ to H$_2$ in an infinite slab exposed to an isotropic external radiation field using brute-force frequency-dependent radiative transfer through the LW bands, coupled to explicit calculation of the H$_2$ level populations. These codes also include a number of physical processes that our treatment omits, such as \hone\ production by cosmic rays, temperature dependence of the H$_2$ formation rate coefficient, and variations in density and temperature through the PDR. By checking our simple analytic model against these much more sophisticated (and computationally costly) calculations, we can estimate the likely level of error in our method.

For this comparison we examine a series of calculations performed by M.~G.\ Wolfire (2009, private communication) using the Meudon PDR code \citep{le-petit06a, kaufman06a}. In these models the radiation field outside the slab is set to $G_0'=0.3$, 0.5, 1.5, 3, 10, 30, or 100, where $G_0'$ is measured in units of the \citet{draine78} field. The dust cross section in the Meudon runs is $\sigma_d=1.1\times 10^{-21}$ cm$^{2}$. The H$_2$ formation rate coefficient varies with temperature, but a rough mean value is $\mathcal{R}=3\times 10^{-17}$ cm$^3$ s$^{-1}$. Most runs use a cosmic ray ionization rate $\zeta=10^{-17}$ s$^{-1}$ (the canonical Milky Way value), but one also uses $\zeta=10^{-16}$ s$^{-1}$, a factor of 10 higher. The gas density in the Meudon code can be set to a constant value, but to provide a more stringent test for how our method fares when there is a realistic level of density variation, we instead compare to runs where the density is determined in one of two self-consistent ways. In some runs the density is determined by requiring that the thermal pressure in every cell have a constant value $P_{\rm th}/k_B = 10^4$ or $10^5$ K cm$^{-3}$; the density and temperature are then obtained iteratively by requiring that the gas be in mechanical and thermal equilibrium at this pressure. In the second group of runs, the mean density as a function of depth in the molecular cloud is determined by requiring that the gas be in hydrostatic equilibrium with the self-gravity of a molecular cloud of a specified mass $M_{\rm cl}$. In these models the median density also differs from the mean density due to the effects of turbulence-induced clumping. For these models the density at a given depth is fully determined by the cloud mass $M_{\rm cl}$, which varies $10^5$ $M_\odot$ to $3\times 10^6$ $M_\odot$. Full details of the models are given in Wolfire, Hollenbach, \& McKee (2009, in preparation).

In order to compare the numerical results to our analytic predictions, we must choose definitions for ambiguous quantities. First, our model is for constant-density gas, but the density is not constant in the numerical runs. Thus we must choose some sort of average density to plug into our analytic formula for $\chi$. This choice is non-trivial because the density rises sharply as gas converts from \hone\ to H$_2$ in the numerical runs, increasing by an order of magnitude between the predominantly atomic and predominantly molecular region. Clearly our predictions will depend on whether we take the gas density to be the atomic density, the molecular density, or some intermediate value. Second, the atomic-to-molecular transition is fairly sharp, but the atomic fraction never falls to zero exactly in the numerical calculations. Thus we must decide how to define the column density or optical depth of the atomic layer.

We consider two schemes for how to define these two quantities. First, following our approach in Paper I, we can use an integral definition of the optical depth of the shielding layer:
\begin{equation}
\label{eq:numerich2}
\tau_{\rm H_2} \equiv \sigma_d \int n_1 \, dz,
\end{equation}
where $n_1$ is the number density of atomic hydrogen.
In the limit of an infinitely sharp transition from atomic to molecular, this clearly gives the correct optical depth for the \hone\ layer, and this definition has the advantage that we need not choose an arbitrary molecular fraction at which to declare that the transition has occurred. Since the \hone\ fraction falls off faster than exponentially, the integral is guaranteed to converge
(neglecting the tiny amount of \hone\ produced by cosmic rays). 
For this definition of $\tau_{\rm H_2}$, since we are sampling the entire gas column where the atomic to molecular transition occurs, we choose to define the density to be the mass-averaged density of all computational cells with $f_{\rm H_2}
<0.9$, where $f_{\rm H_2}\equiv 2n_2/\nh$
 is the fraction of the gas in the molecular phase. Thus our definition takes the density to be the average density considering both the predominantly atomic gas and the gas in the transition zone where the molecular fraction is below 90\%. However, we also considered alternate methods to compute the mean density, ranging from considering only cells with $f_{\rm H_2} < 0.5$ (i.e.\ excluding more of the transition zone) up to allowing all cells with $f_{\rm H_2} < 0.95$ (i.e.\ going further into the molecule-dominated region). These alternate definitions do not change the predicted value of $\tau_{\rm H_2}$ by more than a few percent.

The second option we consider is to define $\tau_{\rm H_2}$ as the dust optical depth from the edge of the slab up to the point where the $f_{\rm H_2} = 0.5$. We refer to this as the point definition, since we are choosing to measure the column up to some specific point. The choice of $0.5$ is rather arbitrary, but picking a particular point allows us to study how big an error we might make by assuming that the atomic-to-molecular transition is sharp, something that an integral definition such as equation (\ref{eq:numerich2}) might obscure. For this point definition of $\tau_{\rm H_2}$, we need to define the density differently than we did using the integral definition -- clearly the density of gas in which $f_{\rm H_2}$ is in the range $0.5-0.9$, which we include in our averaging for the integral definition, cannot affect the propagation of radiation in the region where $f_{\rm H_2} \leq 0.5$. Instead, since the point where $f_{\rm H_2} = 0.5$ is primarily determined by how much radiation is absorbed in the predominantly atomic region, we estimate the density in this case by computing the mass-averaged density of gas in which $f_{\rm H_2} < 0.05$ (i.e.\ in the mainly atomic region). If we instead consider gas with $f_{\rm H_2} < 0.025$ or gas with $f_{\rm H_2} < 0.1$, our predicted optical depths $\tau_{\rm H_2}$ change by $\sim 50\%$.

Now that we have defined a mean density $n_{\rm H}$, we must compute the other quantities that determine $\chi$: the radiation field $E_0^*$ and the H$_2$ formation rate coefficient $\mathcal{R}$. For the radiation field, the photon number density $E_0^*$ corresponding to a radiation intensity $G_0'$ at the cloud surface is $E_0^*= 2 G_0' (7.5\times 10^{-4}\mbox{ cm}^{-3})$, where $7.5\times 10^{-4}$ cm$^{-3}$ is the LW photon density corresponding to the \citet{draine78} field, and the factor of 2 arises because 
the free-space value of the radiation field, $\eosh$, is twice the value at the surface of an infinite slab. 
For the H$_2$ formation rate coefficient, we use $\mathcal{R}=3\times 10^{-17}$ cm$^{3}$ s$^{-1}$, the rough mean value given the temperature variation within the PDR. Finally, we have $\sigma_d=1.1\times 10^{-21}$ cm$^2$ and $f_{\rm diss}=0.11$. Given these values, we compute $\chi$ for each Meudon run, and then we compute the dust optical depth of the \hone\ -- H$_2$ transition predicted by our analytic model using equation (\ref{eq:dustyslab_analyt}).

The circles in the upper panel of Figure \ref{fig:dustyslab} show $\tau_{\rm H_2}$ versus $\chi$ for the integral definition for $\tau_{\rm H_2}$ and density, while the triangles show the results if we instead adopt the point definition. In the range $ \chi\ga 1$, the choice of definition matters little and the errors are small. In this region the largest absolute difference between our analytic model and any of the numerical results is 23\% or 28\% and the median absolute differences are 4\% and 9\% for the integral and point definitions, respectively.  Overall our $\sim 10\%$ error level is comparable to the intrinsic uncertainty in the numerical models arising from our imperfect knowledge of input parameters such as the H$_2$ formation rate coefficient and the dust cross section for LW photons.

In the region $\chi \ll 1$, the two definitions begin to diverge. Our analytic model still tracks the integral definition reasonably well, reaching a maximum error of 40\% at $\chi = 0.078$, the lowest $\chi$ numerical run. In contrast, our model becomes increasingly inaccurate in estimating the location of the 50\% molecular point as $\chi \rightarrow 0$. This is not a surprising result. For $\chi\ll 1$, we found in Paper I and we find again here that the transition between the atomic and molecular regions ceases to be sharp, and a majority of the atomic column is located in regions where the molecular fraction is $>50\%$. One can no longer identify a single meaningful location for the atomic-to-molecular transition. In this case our analytic model remains reasonably accurate in estimating the total atomic column, but, since we do not track the atomic fraction through the transition region, it does not provide a good estimate of where the atomic fraction reaches some specified value. This is analogous to the case of a Str\"omgren analysis for ionizing radiation applied to a situation where the ionization front is not sharp (e.g.\ for ionization by x-rays rather the EUV photons). In such a case there is no well-defined ionization front, and the Str\"omgren analysis does not give a good estimate for where the ionization fraction reaches $50\%$ or some other specified value. However, it does still provide a good estimate for the total emission measure, much as our method still gives a reasonably good estimate for the total \hone\ column.

We emphasize again that in Paper II we showed that $\chi \sim 2-3$ for any cloud, GMC or diffuse, in a region where two-phase equilibrium of the \hone\ gas prevails. Values of $\chi \ll 1$, where our method fails, can only be found in regions where the pressure is either too low (e.g.\ well off the galactic plane, or in the far outskirts of galactic disks) or too high (e.g.\ in the nuclei of starburst galaxies) for two-phase equilibrium to be possible.

\subsection{Spherical Clouds}
\label{sec:dustysphere}

Our algorithm for the spherical case with dust is a slightly more complex version of the ones described in \S~\ref{sec:prisphere} and \S~\ref{sec:dustyslab}. We proceed as follows:
\begin{enumerate}
\item  Select values for $\tau_R$ and $x_{\rm H_2}$. Our algorithm will determine the corresponding $\chi$.
\item Pick a trial value of the parameter $x_f$. (The value of $\phik$ is fixed to $4/3$, as discussed in \S~\ref{sec:prislab}.) The choice of $x_f$ determines the Eddington factor $f_{\rm Edd}$ via equation (\ref{eq:fedd}). 
\item Given $f_{\rm Edd}$, determine the function $\esh$ by integrating the ordinary differential equations (\ref{eq:fsh}) and (\ref{eq:feddesh})
 from $x=x_{\rm H_2}$ to $x=1$, using the boundary condition that $\esh(x_{\rm H_2})=0$.
\item Check whether the solution obeys the integral constraint equation (\ref{eq:constraint}) to within some specified tolerance. If not, repeat steps 2--4 using Newton-Raphson iteration to search for a value of $x_f$ that minimizes the difference between the left- and right-hand sides of equation (\ref{eq:constraint}).
\item If the solution does satisfy the constraint equation (\ref{eq:constraint}), determine $\chi$ from $\esh(1)$ and $\tau_R$ using equation (\ref{eq:eosh}).
\end{enumerate}

\begin{figure}
\plotone{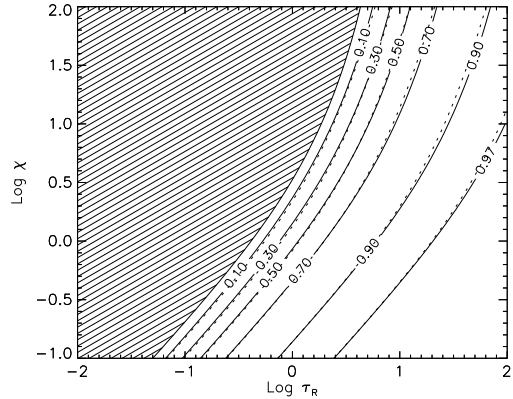}
\caption{
\label{fig:dustsphere_cont}
Contours of $x_{\rm H_2}^3$ as a function of $\tau_R$ and $\chi$. Solid lines show the numerical result, dotted lines show contours computed using the analytic approximation given by equation (\ref{eq:dustsphere_analyt}). The hatched region indicates where $x_{\rm H_2}^3 = 0$.
}
\end{figure}

We use this procedure to calculate values of $\chi$ on a grid of values running from $\tau_R=0.01-100$ and $x_2=0.01-0.99$, and we use this table to construct contours of the molecular volume $x_2^3$ in terms of $\chi$ and $\tau_R$. The results are shown in Figure \ref{fig:dustsphere_cont}. As illustrated in the figure, the numerical result is well-approximated by
\begin{equation}
\label{eq:dustsphere_analyt}
x_{\rm H_2}^3 \simeq 1 - \left(\frac{3}{4}\right) \frac{\xi_d}{1 + 0.0712\,\xi_d^{2.8}},
\end{equation}
where
\beqa
\xi_d &\equiv& \xi\left[\frac{\ln (1 + 0.6\chi + 0.01\chi^2)}{0.6\chi}\right],\\
&=&\frac{71}{\calr_{-16.5} N_{R,21}}\left(\frac{G_0'}{\nh}\right)
\left[\frac{\ln(1+0.6\chi+0.01\chi^2)}{0.6\chi}\right],
\eeqa
and where $N_{R,21}=\nh R/(10^{21}$~cm\ee) is the normalized column density in the cloud;
recall that $\xi\equiv \chi/\tau_R$.
Note that the factor $0.01\chi^2$ is only important to fit the contours when $\chi\gg 1$, and for $\chi\sim 1$, the regime relevant to real molecular clouds, it may be omitted. Also note that equation (\ref{eq:dustsphere_analyt}) is the same as equation (\ref{eq:nodust_analyt}) for the dust-free spherical case, and the only difference between the two cases is that 
$\xi_d$ allows for 
a finite dust optical depth. In the limit $\sigma_d\rightarrow 0$ (so that $\chi\rightarrow 0$), we have $\xi_d\rightarrow \xi$
and this result for $x_{\rm H_2}$ approaches that found in \S \ref{sec:prisphere} above
for dust-free spheres.

Altogether, we have results that cover most of the parameter space in Figure
\ref{fig:dustsphere_cont} except for the region in and immediately adjacent to
the fully atomic region. The results for a dusty sphere obtained in this section
apply for $\chi\ga 1$ and $\tau_R\ga 1$ (the latter requirement is needed to ensure
that the cloud is large enough for the dust to attenuate the radiation).
For the region in which $\chi<1$, the results from \S \ref{sec:prisphere} are valid
provided $\xi=\chi/\tau_R\la 1$; but this is most of the region not adjacent to the fully atomic region. 
According to Paper II, the value of $\chi$ in
galaxies is typically $\sim 2-3$, depending on metallicity (eq. \ref{eq:chigal}).

\section{Molecular Mass Fractions in Atomic-Molecular Complexes}

Thus far we have calculated molecular fractions under the assumption of uniform-density gas. However, it is convenient to generalize to the case where the atomic and molecular regions have different densities, since this is a more realistic representation of an atomic-molecular complex. Given a method to calculate $x_{\rm H_2}$ as a function of $\xi_d$ and $\tau_R$, we solved the problem of determining the molecular mass fraction in such a cloud in Paper II. In that paper we used the approximate form for $x_{\rm H_2}(\chi, \tau_R)$ determined in Paper I; in this section we apply the same approach to calculate the mass fraction for our new, improved determination of $x_{\rm H_2}(\chi, \tau_R)$.

As discussed in Paper II, in a cloud where the atomic and molecular gas have differing densities, determining the molecular mass fraction requires an additional constraint beyond the one imposed by radiative and chemical equilibrium. The most natural constraint is to require that the atomic and molecular gas be in pressure balance. In an atomic-molecular complex of total (atomic plus molecular) surface density $\Sigma_{\rm comp}$ this implies that
\begin{equation}
\label{eq:presbalance}
\tau_c = \tau_R [1 + (\phi_{\rm mol} - 1) x_{\rm H_2}^3],
\end{equation}
where
\begin{equation}
\tau_c \equiv \frac{3}{4} \left(\frac{\Sigma_{\rm comp} \sigma_d}{\mu_{\rm H}}\right),
\end{equation}
$\phi_{\rm mol}\approx 10$ is the ratio of molecular to atomic gas volume densities, and $\mu_{\rm H}$ is the mean mass per H nucleus. Given either an analytic approximation for $x_{\rm H_2}(\chi,\tau_R)$ (e.g.\ equation \ref{eq:dustsphere_analyt}), or a numerical algorithm to calculate $x_{\rm H_2}(\chi,\tau_R)$ (e.g.\ \S~\ref{sec:dustysphere}), equation (\ref{eq:presbalance}) implicitly determines the effective optical depth $\tau_R$ as a function of $\chi$. It is then simple to calculate $\tau_R$ and $x_{\rm H_2}(\chi,\tau_R)$ for a given value of $\chi$, and then to compute the corresponding molecular mass fraction
\begin{equation}
\label{eq:fh2comp}
f_{\rm H_2,\, comp} = \frac{\phi_{\rm mol} x_{\rm H_2}^3}{1 + (\phi_{\rm mol}-1) x_{\rm H_2}^3}.
\end{equation}

Alternately, one can obtain an approximate solution to equation (\ref{eq:presbalance}) analytically.
Equations (\ref{eq:presbalance}) and (\ref{eq:fh2comp}) imply the simple relation
\begin{equation}
\tau_c (1 - f_{\rm H_2,\, comp}) = \tau_R (1 - x_{\rm H_2}^3).
\end{equation}
Solving this for $f_{\rm H_2,\, comp}$ and substituting in our approximate form for $x_{\rm H_2}$ from equation (\ref{eq:dustsphere_analyt}) gives
\begin{eqnarray}
f_{\rm H_2,\, comp} & = & 1 - \frac{\tau_R}{\tau_c} \left(\frac{3}{4}\right) \frac{\xi_d}{1+ 0.0712 \xi_d^{2.8}} \\
& = & 1 - \left(\frac{3}{4}\right) \frac{s}{1+ 0.0712 \xi_d^{2.8}},
\label{eq:fh2comp1}
\end{eqnarray}
where we have defined
\begin{equation}
s \equiv \frac{\tau_R}{\tau_c} \xi_d = \frac{\ln(1 + 0.6 \chi + 0.01 \chi^2)}{0.6 \tau_c}.
\end{equation}
Note that $s$ is defined solely in terms of given parameters. The remaining unknown on the right-hand side of equation (\ref{eq:fh2comp1}) is $0.0712\xi_d^{2.8}$, but since this is a small correction in the denominator, we can approximate it. Re-arranging equation (\ref{eq:presbalance}), we find
\begin{equation}
\xi_d = \frac{\phi_{\rm mol} s}{1 + \left(\frac{3}{4}\right) \frac{(\phi_{\rm mol}-1) s}{1 + 0.0712\xi_d^{2.8}}},
\end{equation}
and experimentation with numerical solutions to this equation shows that $0.0712\xi_d^{2.8}\approx 0.25 s$ for $\phi_{\rm mol} \approx 10$. Substituting this into equation (\ref{eq:fh2comp1}), we arrive at our approximate solution:
\begin{equation}
\label{eq:fh2approx}
f_{\rm H_2,\, comp} \simeq 1 - \left(\frac{3}{4}\right) \frac{s}{1 + 0.25 s}.
\end{equation}
We apply this only for $s<2$; for $s\geq 2$, we have $f_{\rm H_2,\, comp} = 0$.

\begin{figure}
\plotone{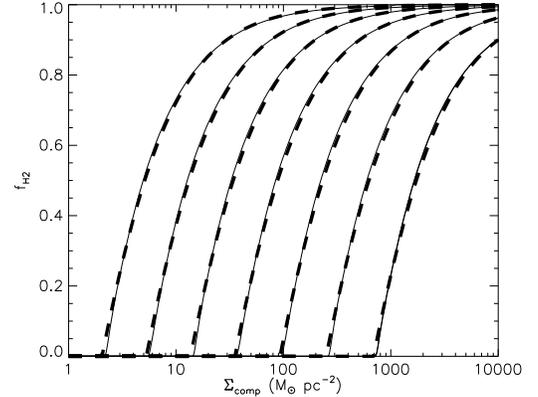}
\caption{
\label{fig:fh2plot}
$f_{\rm H_2}$ vs.\ $\Sigma_{\rm comp}$. Solid lines show numerical solutions of equation (\ref{eq:presbalance}), using the algorithm described in \S~\ref{sec:dustysphere} to compute $x_{\rm H_2}(\chi,\tau_R)$. Dashed lines show the analytic approximation given by equation (\ref{eq:fh2approx}). Pairs of lines correspond to different metallicities, ranging from $\log Z'=-2$ ({\it furthest right}) to $\log Z'=1$ ({\it furthest left}), in intervals of $\log Z'=0.5$.
}
\end{figure}

Following Paper II, we can estimate $\chi$ for a two-phase atomic medium using equation (\ref{eq:chieqn}). Similarly, we can write $\tau_c$ in terms of the complex surface density and the metallicity as $\tau_c = 0.066 \Sigma_{\rm comp,0} Z'$, where $\Sigma_{\rm comp,0} = \Sigma_{\rm comp}/(1\,M_{\odot} \mbox{ pc}^{-2})$ and $Z'$ is the metallicity relative to Solar. In writing this relationship, we have assumed the $\sigma_d \propto \mathcal{R}\propto Z'$. With these estimate for $\chi$ and $\tau_c$, we obtain the relationship between H$_2$ mass fraction and $\Sigma_{\rm comp}$ shown in Figure \ref{fig:fh2plot}. The difference between the approximate solution given by equation (\ref{eq:fh2approx}) and the numerical solution is less than 5\% for all $\Sigma_{\rm comp}$ and metallicities between $\log Z'=-2$ and $\log Z'=1$, where the difference between the approximate and numerical solutions is defined to be $|f_{\rm H_2,approx} - f_{\rm H_2,num}| / \max(f_{\rm H_2,num}, 1-f_{\rm H_2,num})$.

\section{Conclusions}

Using a \strom-type analysis,
we have evaluated the thickness of the \hone\ layer around molecular clouds for
both planar and spherical geometry, and both without and with dust.
We have assumed that the density of the \hone\ layer is constant.
In contrast with most treatments of PDRs, we have assumed that the
radiation incident upon the cloud is isotropic.
The results of our  analysis 
are based on several approximations. The major physical approximation
is that the gas in the layer is entirely atomic, $\fone=1$.
We also assumed that the metallicity $Z\la Z_\odot$, so that the individual absorption
lines in the Lyman-Werner bands overlap and the dissociating
radiation covers most of the band 91.2-110.8 nm \citep{sternberg88}.
 The principal methodological
approximation is the treatment of the variable Eddington factor, $\fedd$, as
described in \S\ref{sec:edd}. 
In the absence of dust, 
$\fedd$ for a spherical cloud has the unusual property that it is generally less than the
isotropic value of $\frac 13$.
 In addition, we assumed that the opacity associated
with the flux was related to that associated with the number density by a constant,
$\phik$. In the one case in which we could determine $\phik$ as an eigenvalue
of the solution (\S\ref{sec:prislab}), we found that this was a good approximation.
Finally, in calculating the optical depth needed to obtain a constraint equation,
we replaced the optical depth for individual rays, $\tau_I$, with the angle-averaged
value, $\tau_E$.
The accuracy of our approximations for the planar case has been tested by comparison
with the results of the Meudon PDR code, which were kindly supplied by M.\ G.\ Wolfire.
The analytic fit to our results (eq. \ref{eq:dustyslab_analyt}) typically agrees with the results
of the PDR code to within 10\%, with the maximum error of about 30\%.

Our results for the scaling behavior of the solutions are consistent with those of
Paper I. In the planar, dust-free case, the solution depends only on
the characteristic \strom\ length, $\lch\propto \eosh/\nh^2 \calr$. When
dust is included, the solution depends only on the dust optical depth associated
with this length scale, $\chi=\nh\sigma_d\lch
\propto G_0'/\nh$. 
In the spherical case,
an additional parameter enters, the dust opacity associated with the cloud
radius: $\tau_R=\nh\sigma_d R$. In the dust-free case, the solution depends
only on the ratio of these two parameters, $\xi=\chi/\tau_R=\lch/R
\propto G_0'/(\nh N_R)$.
We have presented simple analytic approximations for our results that should
be useful for applications.

\acknowledgements
We thank Jason Tumlinson for his contributions to the previous papers in this series and
Mark Wolfire for providing the results from the Meudon PDR code so
that we could test our results. 
We particularly thank
Amiel Sternberg for a number of helpful comments that significantly improved the paper.
Support for this work was provided by  the National Science Foundation through grants 
AST-0606831 (CFM), AST-0908553 (CFM), and
AST-0807739 (MRK), by NASA through the Spitzer  Space Telescope Theoretical Research Program, provided by a contract issued by the Jet Propulsion Laboratory (MRK), and by the Alfred P.\ Sloan Foundation (MRK).

\begin{appendix}
\section{Primordial Formation of \htwo}

In the simplest case, \htwo\ formation in low-density
primordial gas is governed by just two reactions, the formation of H$^-$,
\beq
\rm{H+e^-\rightarrow H^-+h\nu},
\eeq 
and the formation of \htwo,
\beq
\rm{H^- +H\rightarrow H_2 + e^-}.
\eeq
The reaction rates for number of important reactions in primordial gas have been
summarized recently by \citet{
glo08}. The rate coefficient they cite for H$^-$ formation is within
15\% of the simple expression
\beq
k_-=1.83\times 10^{-18}T^{0.88}~~~\mbox{cm$^3$~s\e}
\eeq
given by \citet{hut76}. They indicate that the rate coefficient they give for formation of \htwo,
\beq
k_2=1.3\times 10^{-9}~~~\mbox{cm$^3$~s\e},
\eeq
is uncertain by at least a factor 2. If these two reactions are the only ones governing
the abundance of H$^-$---in particular, if photodetachment by infrared radiation
is negligible---then the equilibrium abundance of H$^-$ is
\beq
n_-=\left(\frac{k_-}{k_2}\right)n_e.
\eeq
The formation rate coefficient $\calr_-$ that is analogous to the rate coefficient
$\calr$ for \htwo\ formation on dust grains,
\beq
\nh^2\calr_-=k_2n_-\nh,
\eeq
is then given by
\beq
\calr_-=8\times 10^{-19}x_{e,\,-3} T_3^{0.88}~~~\mbox{cm$^3$~s\e},
\eeq
where the normalized ionization,
$x_{e,\,-3}\equiv n_e/(10^{-3}\nh)$, is typical
of regions of primordial star formation (e.g., \citealp{abe02,bro02}).

While this analysis shows that in principle our formalism can apply to 
\htwo\ formation in primordial clouds via the H$^-$ process, 
it is not clear whether it can do so in practice.
For temperatures $\sim 10^3$~K, the H$^-$ rate coefficient is about 30 times
less than the rate coefficient for \htwo\ formation in the Milky Way.
As a result, the time scale for \htwo\ formation is very long unless
the density is high, but if the density is high it is likely that the
ionization will be less than the assumed value of $10^{-3}$ and that the formation-rate coefficient
will be correspondingly reduced. For example, in
their calculations of radiative feedback by the first stars, \citet{ahn07}
found that the \htwo\ concentrations were generally less than about 1\%,
whereas our analysis determines the condition for the concentration to approach 100\%.

At higher densities, \htwo\ can form via three-body collisions at a rate
\beq
\nh^2\calr_{3b}=\nh^3 k_{3b},
\eeq
so that the rate coefficient $\calr_{3b}$ is proportional to the density.
The three-body formation rate is uncertain by two orders of magnitude
\citep{glo08};
the geometric mean of the high and low rates they cite is
$k_{3b}=2.4\times 10^{-28}T^{-1.27}$~cm$^6$~s\e\ for $T>300$~K,
corresponding to $\calr_{3b}=3.7\times 10^{-22}(\nh/10^{10}~\mbox{cm\eee})T_3^{-1.27}$ cm$^3$~s\e. Since three-body formation is important only at high densities, high
UV fluxes, such as those in circumstellar disks or in close binaries, are required
to create significant photodissociation zones. Such calculations have yet to be carried out.

\end{appendix}


\begin{thebibliography}

\bibitem[Abel et al.(2002)]{abe02} Abel, T., Bryan, G.~L., 
\& Norman, M.~L.\ 2002, Science, 295, 93 

\bibitem[Ahn 
\& Shapiro(2007)]{ahn07} Ahn, K., \& Shapiro, P.~R.\ 2007, \mnras, 375, 881 

\bibitem[Black \& van Dishoeck(1987)]{black87} Black, J.~H., \& van Dishoeck, E.~F. 1987, \apj, 322, 412

\bibitem[Bromm et al.(2002)]{bro02} Bromm, V., Coppi, P.~S., 
\& Larson, R.~B.\ 2002, \apj, 564, 23 

\bibitem[Browning et al.(2003)]{bro03} Browning, M.~K., 
Tumlinson, J., \& Shull, J.~M.\ 2003, \apj, 582, 810 

\bibitem[Draine(1978)]{draine78} Draine, B.~T. 1978, \apjs, 36, 595

\bibitem[Draine 
\& Bertoldi(1996)]{dra96} Draine, B.~T., \& Bertoldi, F.\ 1996, \apj, 468, 269 

\bibitem[Elmegreen(1993)]{elmegreen93} Elmegreen, B.~G. 1993, \apj, 411, 170

\bibitem[Federman et al.(1979)]{federman79} Federman, S.~R., Glassgold, A.~E., \& Kwan, J. 1979, \apj, 227, 466

\bibitem[Glover 
\& Abel(2008)]{glo08} Glover, S.~C.~O., \& Abel, T.\ 2008, \mnras, 388, 1627 

\bibitem[Gnedin et al.(2009)]{gnedin09} Gnedin, N.~Y., Tassis, K., \& Kravtsov, A.~V. 2009, \apj, 697, 55

\bibitem[Hollenbach 
\& Tielens(1999)]{hol99} Hollenbach, D.~J., \& Tielens, A.~G.~G.~M.\ 1999, Reviews of Modern Physics, 71, 173 

\bibitem[Hutchins(1976)]{hut76} Hutchins, J.~B.\ 1976, \apj, 
205, 103 

\bibitem[Kaufman et al.(2006)]{kaufman06a} Kaufman, M.~J., Wolfire, M.~G., \& Hollenbach, D.~J. 2006, \apj, 644, 283

\bibitem[Krumholz \& McKee(2005)]{kru05} Krumholz, M. R., \&  McKee, C. F. 2005, \apj, 630, 250


\bibitem[Krumholz et al.(2008)]{kru08} Krumholz, M.~R., 
McKee, C.~F., \& Tumlinson, J.\ 2008, \apj, 689, 865 (Paper I)

\bibitem[Krumholz et al.(2009a)]{kru09a} ---. 2009a, \apj, 693, 216 (Paper II)

\bibitem[Krumholz et al.(2009b)]{kru09b} ---. 2009b, \apj, 699, 850

\bibitem[Krumholz et al.(2009c)]{kru09c} Krumholz, M.~R., Ellison, S.~L., Prochaska, J.~X., \& Tumlinson, J. 2009, \apjl, 701, L12

\bibitem[Le Petit et al.(2006)]{le-petit06a} Le Petit, F., Nehm\'e, C., Le Bourlot, J., \& Roueff, E. 2006, \apjs, 164, 506

\bibitem[Liszt(2002)]{liszt02} Liszt, H. 2002, \aap, 389, 393

\bibitem[Liszt \& Lucas(2000)]{liszt00} Liszt, H., \& Lucas, R. 2000, \aap, 355, 333

\bibitem[Neufeld \& Spaans(1996)]{neufeld96} Neufeld, D.~A., \& Spaans, M. 1996, \apj, 473, 894

\bibitem[Petrosian et al.(1972)]{pet72} Petrosian, V., Silk, 
J., \& Field, G.~B.\ 1972, \apjl, 177, L69 

\bibitem[Robertson \& Kravtsov(2008)]{robertson08} Robertson, B.~E., \& Kravtsov, A.~V. 2008, \apj, 680, 1083

\bibitem[Shu(1991)]{shu91} Shu, F.~H.\ 1991, Physics of 
Astrophysics, Vol.~I, University Science 
Books, ISBN 0-935702-64-4, p. 44

\bibitem[Spaans \& Neufeld(1997)]{spaans97} Spaans, M., \& Neufeld, D.~A. 1997, \apj, 484, 785

\bibitem[Sternberg(1988)]{sternberg88} Sternberg, A. 1988, \apj, 332, 400

\bibitem[Stoerzer et 
al.(1996)]{sto96} Stoerzer, H., Stutzki, J., \& Sternberg, A.\ 1996, \aap, 310, 592 



\bibitem[Str{\"o}mgren(1939)]{str39} Str{\"o}mgren, B.\ 1939, 
\apj, 89, 526 

\bibitem[van Dishoeck \& Black(1986)]{vandishoeck86} van Dishoeck, E.~F., \& Black, J.~H. 1986, \apjs, 62, 109

\bibitem[Wolfire et al.(2003)]{wol03} Wolfire, M.~G., McKee, 
C.~F., Hollenbach, D., \& Tielens, A.~G.~G.~M.\ 2003, \apj, 587, 278 


\end{thebibliography}
\end{document}